\documentclass[12pt]{article}

\usepackage{amssymb}
\usepackage{amsmath}

\begin{document}

\title{\bf Virtual gravitons and brane field scattering in the
RS model with a small curvature}

\author{A.V. Kisselev\thanks{Electronic address:
alexandre.kisselev@mail.ihep.ru} \\
\small Institute for High Energy Physics, 142281 Protvino, Russia}

\date{}

\maketitle

\thispagestyle{empty}

\bigskip

\begin{abstract}
The contribution of virtual $s$-channel Kaluza-Klein (KK)
gravitons to high energy scattering of the SM fields in the
Randall-Sundrum (RS) model with two branes is studied. The small
curvature option of the RS model is considered in which the KK
gravitons are narrow low-mass spin-2 resonances. The analytical
tree-level expression for a process-independent gravity part of
the scattering amplitude is derived, accounting for nonzero
graviton widths. It is shown that one cannot get a correct result,
if a series of graviton resonances is replaced by a continuous
mass distribution, in spite of the small graviton mass splitting.
Such a replacement appeared to be justified only in the
trans-Planckian energy region.
\end{abstract}




\section{RS scenario with the small curvature}
\label{sec:RS_model}

The models with spacial extra dimensions (ED's) have pretensions
to solving theoretical problems which have not yet received a
satisfactory answer (such as hierarchy problem, proton lifetime,
hierarchy of fermion masses and mixing angles, \emph{etc}) and
lead to a new phenomenology in the TeV energy region. The
multidimensional gravity is strong, and the fundamental Planck
scale can be related with the string scale. One manifestation of
theories with ED's is the existence of Kaluza-Klein (KK) gravitons
and their interactions with the SM fields.

In the present paper, we consider one realization of the ED theory
in a slice of the AdS$_5$ space-time with the following background
warped metric:
\begin{equation}\label{metric}
ds^2 = e^{2 \kappa (\pi r - |y|)} \, \eta_{\mu \nu} \, dx^{\mu} \,
dx^{\nu} + dy^2,
\end{equation}
where $y = r \theta$ ($-\pi \leqslant \theta \leqslant \pi$), $r$
being the ``radius'' of extra dimension, and  $\eta_{\mu \nu}$ is
the Minkowski metric. The parameter $\kappa$ defines a
5-dimensional scalar curvature of the AdS$_5$ space.

We will be interested in the Randall-Sundrum (RS)
model~\cite{Randall:99} which has two 3D branes with equal and
opposite tensions located at the point $y = \pi r$ (called the
\emph{TeV brane}, or \emph{visible brane}) and point $y = 0$
(referred to as the \emph{Plank brane}). If $k > 0$, then the
tension on the TeV brane is negative, whereas the tension on the
Planck brane is positive. All the SM fields are constrained to the
TeV brane, while the gravity propagates in five dimensions.

The main goal of the paper is to estimate $s$-channel graviton
contribution to the scattering of the \emph{brane} fields in such
a scheme.

Let us note that the warp factor in the metric~\eqref{metric} is
equal to 1 on the negative tension (visible) brane, and a correct
determination of particle masses on this brane is thus
achieved~\cite{Rubakov:01}. By calculating the zero mode sector of
the effective theory, one then obtains the ``hierarchy relation'',
\begin{equation}\label{RS_hierarchy_relation}
\bar{M}_{\mathrm{Pl}}^2 = \frac{\bar{M}_5^3}{\kappa} \left( e^{2
\pi \kappa r} - 1 \right),
\end{equation}
with $\bar{M}_5$ being a 5-dimensional Planck scale.

From the point of view of an observer located on the TeV brane,
there exists an infinite number of graviton KK excitations with
masses
\begin{equation}\label{graviton_masses}
m_n = x_n \, \kappa, \qquad n=1,2 \ldots ,
\end{equation}
where $x_n$ are zeros of the Bessel function $J_1(x)$, with
\begin{equation}\label{zeros}
x_n = \pi \, \Big( n + \frac{1}{4} \Big) + \mathrm{O \left( n^{-1}
\right)} .
\end{equation}
Note that \emph{all} zeros of $J_{1}(x)/x$ are simple ones, and
that they are \emph{real positive} numbers~\cite{Watson}.%
\footnote{The minimal positive zero of the Bessel function
$J_{\nu}(z)$ is approximately equal to $\nu + 1.86 \,
\nu^{1/3}$~\cite{Watson}.}

The interaction Lagrangian on the TeV brane looks like (with the
radion field omitted)
\begin{equation}\label{Lagrangian}
\mathcal{L} = - \frac{1}{\bar{M}_{\mathrm{Pl}}} \, T^{\mu \nu} \,
G^{(0)}_{\mu \nu} - \frac{1}{\Lambda_{\pi}} \, T^{\mu \nu} \,
\sum_{n=1}^{\infty} G^{(n)}_{\mu \nu}.
\end{equation}
Here $T^{\mu \nu}$ is the energy-momentum tensor of the matter on
the brane, $G^{(n)}_{\mu \nu}$ is the graviton field with the
KK-number $n$, and
\begin{equation}\label{lambda}
\Lambda_{\pi} = \bar{M}_5 \left( \frac{\bar{M}_5}{\kappa}
\right)^{\! 1/2}
\end{equation}
is a physical scale on the TeV brane (here and in what follows, we
neglect small corrections $\mathrm{O}(e^{-\pi k r})$).

To get $m_1 \sim 1$ TeV, the parameters of the model are usually
taken to be $\kappa \sim \bar{M}_5 \sim 1$ TeV. Then one obtains a
series of massive graviton resonances in the TeV region which
interact rather strongly with the SM particles, since
$\Lambda_{\pi} \sim 1$ TeV.

In the present paper we will consider a different scenario which
we call \emph{\textbf{``small curvature
option''}}~\cite{Giudice:04,Kisselev:05}:
\begin{equation}\label{scale_relation}
\kappa \ll \bar{M}_5 \sim 1 \mathrm{\ TeV}.
\end{equation}
In such a scheme, the scale $\Lambda_{\pi}$ appears to be
significantly larger than the gravity scale $M_5$, as one can see
from Eq.~\eqref{lambda}. Namely, we get $\Lambda_{\pi} = 100 \,
(M_5/\mathrm{TeV})^{3/2}(\mathrm{100 \ MeV}/\kappa)^{1/2}$ TeV.
Nevertheless, for both real and virtual graviton production, a
magnitude of a scattering amplitude is defined by the
5-dimensional Planck scale $\bar{M}_5$, not by $\Lambda_{\pi}$ or
$\kappa$, separately (see formulas in the next Section).%
\footnote{Contrary to the KK gravitons, the radion will be hardly
produced, since the radion coupling to SM fields is equal to
$1/(\sqrt{3} \Lambda_{\pi})$. The radion-Higgs mixing (if it
exists, see \cite{Giudice:01}) will be also suppressed.}

Contrary to the case $\kappa \sim \bar{M}_5 \sim 1$ TeV, there
exists a series of very narrow low-mass spin-2 resonances with an
almost continuous mass distribution. For such a case, the
following inequalities were derived in
Ref.~\cite{Kisselev:05}:%
\footnote{For the case $\kappa \sim \bar{M}_5 \sim
\bar{M}_{\mathrm{Pl}}$, analogous bounds look like $0.01 \leqslant
\kappa/\bar{M}_{\mathrm{Pl}} \leqslant 0.1$~\cite{Davoudiasl:01}.}
\begin{equation}\label{curvature_scale_ratio}
10^{-5} \leqslant \frac{\kappa}{\bar{M}_5} \leqslant 0.1 .
\end{equation}
Notice, in order the hierarchy relation for the warped
metric~\eqref{RS_hierarchy_relation} to be satisfied, we have to
put $\kappa r \approx 10$.

It is worth to underline that the AdS$_5$ space-time differs
significantly from a 5-dimensional flat space-time with one
\emph{large} ED even for very small $\kappa$ (i.e. for the small
curvature). Indeed, let us consider the hierarchy relation for $d$
compact ED's of the size $R_c$:
\begin{equation}\label{ADD_hierarchy_relation}
\bar{M}_{\mathrm{Pl}}^2 = (2\pi R_c)^d \, \bar{M}_{4+d}^{2+d},
\end{equation}
where $ \bar{M}_{4+d}$ is a gravity scale in the flat space-time
with $d$ compact ED's. For $d = 1$, equation
\eqref{ADD_hierarchy_relation} is a particular case of
\eqref{RS_hierarchy_relation} in the limit $2\pi \kappa r \ll 1$.
However, the condition $2\pi \kappa r \ll 1$ means that the ratio
$\bar{M}_5/\kappa$ should be unrealistically large:
\begin{equation}\label{gravity_scale_over_curvature}
\frac{\bar{M}_5}{\kappa} \gg \left(
\frac{\bar{M}_{\mathrm{Pl}}}{\bar{M}_5} \right)^2.
\end{equation}
This inequality means, for example, that $\kappa \ll 10^{-22}$ eV,
if $\bar{M}_5 = 1$ TeV.

The present astrophysical bounds~\cite{Hanhart:01} rule out the
possibility $d = 1$ and
significantly restrict the parameter space for $d = 2, 3$.%
\footnote{If one insists that $\bar{M}_{4+1}$ should be of order
of few TeV, the case $d=1$ is completely excluded, since $R_c$
exceeds the size of the solar system.}
The most stringent constraints come from neutron-star (NS) excess
heat due to the trapped cloud of the KK gravitons surrounding the
NS (for details, see the second paper in \cite{Hanhart:01}). For
instance, one gets $R_c^{-1} > 4.4 \cdot 10^{-12}$ GeV (and,
correspondingly, $\bar{M}_{4+1} > 1.6 \cdot 10^5$ TeV) for $d=1$.

Fortunately, the above mentioned restrictions can not be directly
applied to the AdS$_5$ space-time, since they were derived in the
soft radiation approximation, $\omega \ll T$, where $\omega$ is
the graviton energy, while $T$ is the temperature of the nuclear
medium (for instance, in the NS). Equation
\eqref{curvature_scale_ratio} means that $\omega \geqslant m_1 =
38(\bar{M}_5/\mathrm{TeV})$ MeV. On the other hand, a typical
value of $T$ is equal to 30 MeV. Thus, for $\bar{M}_5 \gtrsim 1$
TeV the condition $\omega \ll T$ is not satisfied, even for
$\kappa = 10 (\bar{M}_5/\mathrm{TeV})$ MeV.

\section{Virtual $\mathbf{s}$-channel gravitons}
\label{sec:virtual_graviton}

Let us study the scattering of two SM fields mediated by massive
graviton exchanges in the $s$-channel,
\begin{equation}\label{process}
a \, \bar{a} \rightarrow G^{(n)} \rightarrow b \, \bar{b} \; ,
\end{equation}
where $a (b) = e^-, \  \gamma, \ q, \ g$, \emph{etc}. For
instance, in hadron collisions virtual graviton effects could be
seen in the processes $pp \rightarrow 2 \mathrm{\ jets} + X$, $pp
\rightarrow \gamma \gamma + X$, and Drell-Yan process $pp
\rightarrow l^+l^- + X$. At linear colliders, the promising
reactions are $e^+e^- \rightarrow \gamma \gamma$ and $e^+e^-
\rightarrow f \bar{f}$. In what follows, the invariant
energy of the process, $\sqrt{s}$, is assumed to be around 1 TeV.%
\footnote{The energy region $\sqrt{s} \gg 1$ TeV will be also
briefly considered, see Eqs.~\eqref{42}-\eqref{48}.}
It means that we are working in the following region:
\begin{equation}\label{energy_region}
 \Lambda_{\pi} \gg \sqrt{s} \sim M_5 \gg \kappa.
\end{equation}

The matrix element of the process~\eqref{process} looks like
\begin{equation}\label{02}
\mathcal{M} = \mathcal{A} \, \mathcal{S} \; .
\end{equation}
The fist factor in Eq.~\eqref{02} contains the following
contraction of tensors:
\begin{equation}\label{03}
\mathcal{A} = T_{\mu \nu}^a \, P^{\mu \nu \alpha \beta} \,
T_{\alpha \beta}^b = T_{\mu \nu}^a \, T^{b \, \mu \nu} -
\frac{1}{3} \, {(T^a)}^{\mu}_{\mu} \, {(T^b)}^{\nu}_{\nu} \; ,
\end{equation}
where $P^{\mu \nu \alpha \beta}$ is a tensor part of the graviton
propagator, while $T_{\mu \nu}^{a(b)}$ is the energy-momentum
tensor of the field $a(b)$.

\subsection{Nonzero graviton widths}
\label{subsec:nonzero_width}

We will concentrate on the second factor in Eq.~\eqref{02} which
is universal for all types of processes mediated by the
$s$-channel exchanges of the KK gravitons. It is of the form:
\begin{equation}\label{04}
\mathcal{S}(s) =  \frac{1}{\Lambda_{\pi}^2} \sum_{n=1}^{\infty}
\frac{1}{s - m_n^2 + i \, m_n \Gamma_n} \; .
\end{equation}
Here $\Gamma_n$ denotes the total width of the graviton with the
KK number $n$ and mass $m_n$.

This sum was estimated in Ref.~\cite{Giudice:04} in a \emph{zero
width approximation} (assuming $\Gamma_n = 0$ for all $n$). It was
shown that $\mathcal{S}(s)$ is purely imaginary in this limit (if
no ultraviolet (UV) cutoff is imposed). Thus, there is no
interference of ED effects with SM contributions to the same
processes.

The width of the massive graviton is indeed very small if its
KK-number $n$ is not too large~\cite{Kisselev:05_2}:
\begin{equation}\label{06}
\frac{\Gamma_n}{m_n} = \eta \left( \frac{m_n}{\Lambda_{\pi}}
\right)^2 ,
\end{equation}
where $\eta \simeq 0.09$.%
\footnote{To estimate $\Gamma_n$, masses of the SM particles were
neglected with respect to the graviton mass
$m_n$~\cite{Kisselev:05_2}. This approximation is well justified
since the sum in $n$ \eqref{04} is mainly given by the gravitons
with the masses $m_n \sim \sqrt{s} \sim 1$ TeV.}
However, the main contribution to the sum~\eqref{04} comes from
the region $n \sim \sqrt{s}/\kappa \gg 1$. So, nonzero widths of
the gravitons in the RS model with the small curvature should be
taken into account. That is why, we will study a general case
($\Gamma_n \neq 0$). For comparison, a particular case (all
$\Gamma_n = 0$) will be also analyzed (see
subsection~\ref{subsec:zero_width}).

It is useful to present $\mathcal{S}(s)$ in the form:
\begin{equation}\label{08}
\mathcal{S}(s) =  \sum_{n=1}^{\infty} \frac{1}{a \, x_n^4 - b \,
x_n^2 + c}  =  \frac{1}{a (\sigma^2 - \rho^2)} \,
\sum_{n=1}^{\infty} \left[ \frac{1}{x_n^2 - \sigma^2} -
\frac{1}{x_n^2 - \rho^2} \right],
\end{equation}
with
\begin{equation}\label{10}
a = i \, \eta \kappa^4, \quad b = (\kappa \Lambda_{\pi})^2, \quad
c = s \Lambda_{\pi}^2 \; .
\end{equation}
Here
\begin{align}
\sigma^2 &= \frac{s}{\kappa^2} \,  \frac{2}{1 + \sqrt{1 -
\displaystyle  4 i \eta \frac{s}{\Lambda_{\pi}^2}}}
\label{12} \\
\intertext{and} \rho^2 &= \frac{1}{2i \eta} \left(
\frac{\Lambda_{\pi}}{\kappa} \right)^2 \left[ 1 + \sqrt{1 - 4 i
\eta \frac{s}{\Lambda_{\pi}^2}} \right] \label{13}
\end{align}
are zeros of the quadratic equation $a \, z^2 - b \, z + c = 0$.
In the kinematical region \eqref{energy_region}, the parameter
$\sigma$ \eqref{12} can be approximated as
\begin{equation}\label{14}
\sigma \simeq  \frac{\sqrt{s}}{\kappa} + \frac{i \eta}{2} \,
\left( \frac{\sqrt{s}}{\bar{M}_5}\right)^3 ,
\end{equation}
with $|\,\sigma| \gg 1$.

The sum in Eq.~\eqref{08} can be calculated analytically by the
use of the formula~\cite{Watson}
\begin{equation}\label{16}
\sum_{n=1}^{\infty} \frac{1}{ z_{n, \nu}^2 - z^2} = \frac{1}{2 z}
\, \frac{J_{\nu + 1}(z)}{J_{\nu}(z)} \; ,
\end{equation}
where $z_{n, \, \nu}$ ($n=1,2 \ldots$) are zeros of the function
$z^{-\nu} J_{\nu}(z)$. As a result, we obtain:

\begin{equation}\label{18}
\mathcal{S}(s) = - \frac{1}{2\kappa \bar{M}_5^3} \,
\frac{1}{\sqrt{1 - \displaystyle 4 i \, \eta
\frac{s}{\Lambda_{\pi}^2}}} \left[ \frac{1}{\sigma} \,
\frac{J_2(\sigma)}{J_1(\sigma)} - \frac{1}{\rho} \,
\frac{J_2(\rho)}{J_1(\rho)} \right].
\end{equation}

Since $\eta s/\Lambda_{\pi}^2 \ll 1$, we have the inequality:
\begin{equation}\label{20}
|\,\rho| \simeq \frac{1}{\sqrt{\eta}} \,
\frac{\Lambda_{\pi}}{\kappa} \gg |\,\sigma| \; ,
\end{equation}
and Eq.~\eqref{18} becomes
\begin{equation}\label{22}
\mathcal{S}(s) = - \frac{1}{2 \kappa \bar{M}_5^3} \,
\frac{1}{\sigma} \, \frac{J_2(\sigma)}{J_1(\sigma)} \; ,
\end{equation}
with $\sigma$ given by Eq.~\eqref{14} (here and in what follows,
small corrections like $\mathrm{O(\kappa/\sqrt{s})}$ are omitted).
The function $\mathcal{S}(s)$ has no singularities, as all zeros
of $J_1(z)$ are real, and $\mathrm{Im} \, \sigma \neq 0$ at
physical $s$.

By using asymptotic behavior of the Bessel function~\cite{Watson}
and formulas \eqref{A10} in Appendix~A, we obtain from \eqref{22}:
\begin{equation}\label{24}
\mathcal{S}(s) = - \frac{1}{4 \bar{M}_5^3 \sqrt{s}} \; \frac{\sin
2A  + i \sinh 2\varepsilon }{\cos^2 \! A + \sinh^2 \! \varepsilon
} \; ,
\end{equation}
where
\begin{equation}\label{26}
A = \frac{\sqrt{s}}{\kappa} + \frac{\pi}{4}, \qquad \varepsilon  =
\frac{\eta}{2} \Big( \frac{\sqrt{s}}{\bar{M}_5} \Big)^3 \; .
\end{equation}
Formulas \eqref{22} and \eqref{24} are \emph{\textbf{our main
result}}.

The following inequalities immediately result from \eqref{24}:
\begin{equation}\label{28}
- \coth \varepsilon  \leqslant \mathrm{Im}
\,\mathcal{\tilde{S}}(s) \leqslant - \tanh \varepsilon  \;
\end{equation}
\begin{align}
\big| \mathrm{Re} \,\mathcal{\tilde{S}}(s) \big| &\leqslant
\frac{1}{1 + 2\sinh^2
\! \varepsilon } \; ,
\label{30}  \\
\left| \frac{\mathrm{Re} \, \mathcal{\tilde{S}}(s)}{\mathrm{Im}
\,\mathcal{\tilde{S}}(s)} \right| &\leqslant \frac{1}{\sinh 2
\varepsilon } \; ,
\label{32}
\end{align}
where the notation $ \mathcal{\tilde{S}}(s) = [2 \bar{M}_5^3
\sqrt{s} \,] \, \mathcal{S}(s)$ is introduced. Note that the upper
bound for the ratio $|\, \mathrm{Re} \,\mathcal{S}(s)/\mathrm{Im}
\,\mathcal{S}(s)|$ decreases rapidly with energy, and it becomes
as small as 0.08 at $\sqrt{s} = 3 \bar{M}_5$. For comparison, this
bound is equal to 0.85 at $\sqrt{s} = \bar{M}_5$. The absolute
value of $\mathrm{Im} \,\mathcal{\tilde{S}}(s)$ tends to 1 when
$s$ grows. For instance, we find $0.98 \leqslant | \, \mathrm{Im}
\,\mathcal{\tilde{S}}(s)| \leqslant 1.02$ at $\sqrt{s} = 3
\bar{M}_5$.

If $\sqrt{s} \leqslant \bar{M}_5$, the parameter $\varepsilon$ is
numerically small, and we obtain that
\begin{equation}\label{34}
\mathcal{S}(s) = - \frac{i}{\eta s^2} \left[ 1 +
\frac{\varepsilon^2}{3} + \mathrm{O}(\varepsilon^4) \right]
\end{equation}
at $\sqrt{s} = z_0 \kappa$, where $z_0$ is some zero of the Bessel
function $J_1(z)$. Thus, the value of $\mathcal{S} (s)$ at the
point $\sqrt{s} = z_0 \kappa$ is actually defined by the graviton
with the mass $m_n = \sqrt{s}$, while the relative contributions
from other KK gravitons are suppressed at least by the factor
$\eta^2/12 \approx 7 \cdot 10^{-4}$.

At ultra-high energies, namely, at
\begin{equation}\label{42}
\sqrt{\eta s} \gg \Lambda_{\pi} ,
\end{equation}
our sum can be approximated as follows:
\begin{equation}\label{44}
\mathcal{S}(s) = \sum_{n=1}^{\infty} \frac{1}{a \, x_n^4 + c} \; ,
\end{equation}
with the parameters $a$ and $c$ being defined above~\eqref{10}.
The sum~\eqref{44} has the following asymptotics (see Appendix~A):
\begin{equation}\label{46}
\mathcal{S}(s) \Big|_{\sqrt{\eta s} \gg \Lambda_{\pi}} \simeq
e^{-i B} \, \frac{1}{2\sqrt{2}} \, \frac{1}{\bar{M}_5^3 \sqrt{s}}
\, \left( \frac{\Lambda_{\pi}^2}{\eta s} \right)^{1/4} ,
\end{equation}
where
\begin{equation}\label{48}
B = 2 \sqrt{2} \, \left( \sin \frac{\pi}{8} \right)
\frac{1}{\kappa} \left( \frac{s \Lambda_{\pi}^2}{\eta}
\right)^{1/4} \! + \frac{\pi}{8} \; .
\end{equation}
Notice, the condition $\sqrt{\eta s} \gg \Lambda_{\pi}$ means that
we are working in the energy region $\sqrt{s} \gg 100 \,
(M_5/\mathrm{TeV})^{3/2} (\mathrm{100 \ MeV}/\kappa)^{1/2}$ TeV.

\subsection{Continuous mass spectrum of the gravitons}
\label{subsec:mass_spectrum}

Since $x_n \simeq \pi n$ at large $n$,%
\footnote{To be more correct, one should use $\frac{dx_n}{dn}
\simeq \pi \left[ 1 + \frac{3}{8(\pi n)^2} \right]$. However, we
can put $x_n = \pi n$ at large values of $n$ which are relevant
for our calculations.}
the mass splitting, $\Delta m_{\mathrm{KK}} \simeq \pi \kappa$, is
very small with respect to the energy, and it seems reasonable to
approximate a summation in $n$ by integration over graviton mass
$m_{\mathrm{KK}}$.

Let us calculate the real and imaginary parts of $\mathcal{S}(s)$
separately and compare the results of these calculations with
Eq.~\eqref{24}. We start from the following expressions:
\begin{align}
\mathrm{Re} \, \mathcal{S}(s) &= \frac{1}{(\kappa
\Lambda_{\pi})^2} \sum_{n=1}^{\infty} \frac{\alpha -
x_n^2}{(\alpha - x_n^2)^2 + \beta x_n^8} \; ,
\label{50} \\
\mathrm{Im} \, \mathcal{S}(s) &= - \frac{\eta}{\Lambda_{\pi}^4}
\sum_{n=1}^{\infty} \frac{x_n^4}{(\alpha - x_n^2)^2 + \beta x_n^8}
\; ,
\label{52}
\end{align}
with $\alpha = s/\kappa^2$ and $\beta = \eta^2
(\kappa/\Lambda_{\pi})^4$. The usual way of calculating
Eqs.~\eqref{50}, \eqref{52} is to replace them by the integrals:
\begin{align}
\mathrm{Re} \, \mathcal{S}(s) &= \frac{1}{\pi} \,\frac{1}{\sqrt{s}
\bar{M}_5^3} \int\limits_{(\pi \kappa)/\sqrt{s}}^{\infty} \!\!\!
dz \, \frac{1 - z^2}{(1 - z^2)^2 + \delta z^8} \; ,
\label{54} \\
\mathrm{Im} \, \mathcal{S}(s) &= - \frac{\eta }{\pi} \,
\frac{\sqrt{s}}{\kappa \Lambda_{\pi}^4} \int\limits_{(\pi
\kappa)/\sqrt{s}}^{\infty} \!\!\! dz \, \frac{z^4}{(1 - z^2)^2 +
\delta z^8} \; ,
\label{56}
\end{align}
where
\begin{equation}\label{58}
\delta = \left( \frac{\sqrt{\eta s}}{\Lambda_{\pi}} \right)^4 \ll
1 .
\end{equation}

These integrals are estimated in Appendix~B. The result of the
calculations is the following:
\begin{align}
\mathrm{Re} \, \mathcal{S}(s) &\simeq \frac{1}{2 \bar{M}_5^3
\sqrt{s}} \left[ \sqrt{\frac{\eta \kappa s}{2 \bar{M}_5^3}} -
\frac{2\kappa}{\sqrt{s}}\right] ,
\label{60} \\
\mathrm{Im} \, \mathcal{S}(s) &\simeq - \frac{1}{2 \bar{M}_5^3
\sqrt{s}} \; .
\label{62}
\end{align}
It follows from Eqs.~\eqref{60}, \eqref{62} that the imaginary
part dominates the real one.

These expressions for the real and imaginary parts of
$\mathcal{S}(s)$ \emph{do not agree} with the ``discrete mass''
expression~\eqref{24}, although they obey inequalities
\eqref{28}-\eqref{32}. Remember that Eq.~\eqref{24} was derived by
the direct calculation of the input sum~\eqref{04}. However, the
imaginary parts become practically the same in both cases in the
trans-Planckian kinematical region, namely, at $\sqrt{s} > 3
\bar{M}_5$, as one can see from \eqref{28} and \eqref{62}. As for
the real parts, they are small in comparison with the imaginary
parts in this energy region.%
\footnote{See the comments and numerical estimates after
Eqs.~\eqref{28}-\eqref{32}.}

Can we approximate the discrete spectrum by the continuous mass
distribution, if the mass splitting $\Delta m_{\mathrm{KK}}$ is
``very small''? First of all, let us note that $\Delta
m_{\mathrm{KK}}$ is dimensional and it should be compared with
another dimensional quantity. Actually, we may regard a set of
narrow graviton resonances to be a continuous mass spectrum
(within some interval of $n$), if only
\begin{equation}\label{continuous_spectrum}
\Delta m_{\mathrm{KK}} < \Gamma_n \;
\end{equation}
is satisfied. Let us stress, it is the inequality that allows one
\emph{to replace a summation} in KK number $n$ \emph{by
integration} over
graviton mass $m_{\mathrm{KK}}$.%
\footnote{From the point of view of experimental measurements, the
mass splitting must be compared with the experimental resolution
$\Delta m_{\mathrm{res}}$. The spectrum looks continuous when
$\Delta m_{\mathrm{KK}} < \Delta m_{\mathrm{res}}$, irrespective
of Eq.~\eqref{continuous_spectrum}.}

In our case, the relevant values of $n$, which give the leading
contribution to the sought for quantity $\mathcal{S}(s)$, are $n
\sim \sqrt{s}/(\pi \kappa)$. Then we obtain from
\eqref{continuous_spectrum} and \eqref{06}:
\begin{equation}\label{64}
\eta \, \frac{(\sqrt{s})^3}{\Lambda_{\pi}^2} > \pi \kappa \; ,
\end{equation}
or, equivalently,
\begin{equation}\label{66}
\sqrt{s} \gtrsim 3 \bar{M}_5 \; .
\end{equation}

It is a common belief that in the flat space-time with large ED's
of the size $R_c$, the mass splitting is so small ($\Delta
m_{\mathrm{KK}} = R_c^{-1}$) that the continuous mass approximation
is undoubtedly valid.%
\footnote{For instance, $R_c^{-1} \approx 130$ eV, for $d=4$ and
$\bar{M}_{4+4} = 1$ TeV.}
Surprisingly, \emph{it is not a case}. The reason is that the
gravitons are extremely narrow resonances, $\Gamma_n \sim
m_n^3/\bar{M_{\mathrm{Pl}}^2}$. Accounting for the hierarchy
relation for $d$ compact ED's \eqref{ADD_hierarchy_relation}, one
finds from \eqref{continuous_spectrum} that only KK gravitons with
unrealistically large masses,
\begin{equation}\label{KK_number}
m_n^3 > \bar{M}_{\mathrm{Pl}}^{2 - \frac{2}{d}} \,
\bar{M}_{4+d}^{1 + \frac{2}{d}} \; ,
\end{equation}
are continuously distributed for $d \geqslant 2$. For all that,
the widths of these gravitons remain relatively small, $\Gamma_n
\gtrsim \bar{M}_{4+d} \,
(\bar{M}_{4+d}/\bar{M}_{\mathrm{Pl}})^{2/d}$.

In particular, we get the conditions (for $m_n \sim \sqrt{s}$):
\begin{alignat}{2}\label{68}
\sqrt{s} & >  \bar{M}_{4+1} \; , & \qquad \mathrm{for \ } & d=1,
\nonumber \\
\sqrt{s} & > (\bar{M}_{\mathrm{Pl}} \, \bar{M}_{4+2}^2)^{1/3}, &
\qquad \mathrm{for \ } & d=2,
\nonumber \\
\sqrt{s} & > (\bar{M}_{\mathrm{Pl}}^2 \, \bar{M}_{4+d})^{1/3}, &
\qquad \mathrm{for \ } & d \gg 1.
\end{alignat}
It is not surprising that the first of these inequalities is
similar to Eq.~\eqref{66}.

\subsection{Zero width approximation}
\label{subsec:zero_width}

Now let us consider the limiting case of zero graviton widths
(stable KK gravitons). The corresponding formulas can be obtained
from the formulas derived above, if one \emph{formally} takes the
limit $\eta \rightarrow 0$, $\eta > 0$ in them (remember that all
$\Gamma_n$ are proportional to $\eta$). Let us introduce the
notation
\begin{equation}\label{70}
\mathcal{S}_0(s) = \mathcal{S} (s) \Big|_{\Gamma_n =0}.
\end{equation}
Then we get from \eqref{24}:
\begin{equation}\label{72}
\mathcal{S}_0(s) = \frac{1}{2 \bar{M}_5^3 \sqrt{s}} \, \left\{
\mathcal{P} \cot \Big( \frac{\sqrt{s}}{\kappa} - \frac{\pi}{4}
\Big) - i \pi \sum_{n=0}^{\infty} \delta \left[
\frac{\sqrt{s}}{\kappa} - \pi \Big( n + \frac{1}{4} \Big) \right]
\right\},
\end{equation}
where $\mathcal{P}$ means the principal value. As one can see,
neither $\mathcal{S}_0(s)$, nor $\mathcal{S}(s)|_{\sqrt{s}=\kappa
z_0}$ \eqref{34} depend on the large mass scale $\Lambda_{\pi}$.

This result~\eqref{72} may be also obtained by direct calculations
in the zero width approximation, if one uses the asymptotics of
$x_n$ at large values of $n$~\eqref{zeros} which are relevant in
our case. Indeed, at $\sqrt{s}/\kappa \gg 1$ the quantity
$\mathcal{S}_0(s)$ is approximated as
\begin{equation}\label{74}
\mathcal{S}_0(s) \Big|_{\sqrt{s}/\pi \kappa \gg 1} =
\frac{1}{\pi^2 \kappa \bar{M}_5^3} \, \sum_{n=1}^{\infty}
\frac{1}{\displaystyle \frac{s}{(\pi \kappa)^2} - \Big( n +
\frac{1}{4} \Big)^2 + i0} \; ,
\end{equation}
and equation \eqref{72} is then reproduced, as it is shown in
Appendix~C.

If we replace the summation in $n$ by integration, we find that
the imaginary part of $\mathcal{S}_0(s)$,
\begin{equation}\label{76}
\mathrm{Im} \, \mathcal{S}_0(s) = - \frac{\pi \kappa}{2
\bar{M}_5^3 \sqrt{s}} \, \sum_{n=1}^{\infty} \delta (\sqrt{s} -
m_n) \; ,
\end{equation}
is of the form:
\begin{equation}\label{78}
\mathrm{Im} \, \mathcal{S}_0(s) = - \frac{1}{2 \bar{M}_5^3
\sqrt{s}} \int\limits_{m_0}^{\infty} \!\! dm \, \delta (\sqrt{s} -
m) = - \frac{1}{2 \bar{M}_5^3 \sqrt{s}} \; ,
\end{equation}
where $m_0 = \pi \kappa$. This expression coincides with
formula~\eqref{62} derived in the previous subsection.

The same procedure, when applied to calculating $\mathrm{Re}
\mathcal{S}_0(s)$, results in
\begin{equation}\label{80}
\mathrm{Re} \, \mathcal{S}_0(s)  = \frac{1}{\pi \bar{M}_5^3 } \,
\mathcal{P} \!\! \int\limits_{m_0}^{\infty} \!\! dm \,
\frac{1}{\displaystyle s - m^2} = - \frac{\kappa}{s \bar{M}_5^3}.
\end{equation}
The last term in \eqref{80} is a particular case of
expression~\eqref{60} in the limit $\eta \rightarrow 0$. As one
can see, $\mathcal{S}_0(s)$ is actually purely imaginary.%
\footnote{The real part is suppressed with respect to the
imaginary part by the factor $\kappa/(2\sqrt{s})$.}
Note that ``continuous mass spectrum'' formulas \eqref{78},
\eqref{80} \emph{are in disagreement} with the imaginary and real
parts of the ``discrete mass spectrum'' expression~\eqref{72}
taken at  $\eta \rightarrow 0$ (this discrepancy  is discussed in
the end of subsection~\ref{subsec:mass_spectrum}).

Let us stress that a more rapid falloff of $\mathcal{S}(s)$ at
$\sqrt{\eta s} \gg \Lambda_{\pi}$ (see Eq.~\eqref{46}) \emph{is
completely lost} in the zero width approximation.

\section{Conclusions and discussions}

In the present paper we have estimated the contribution of the
virtual $s$-channel KK gravitons to the scattering of two SM
fields. We have considered the small curvature option of the RS
model with two branes ($\kappa \ll \bar{M}_5$). In such a scheme,
the KK graviton spectrum is a series of rather narrow low-mass
resonances. All the SM fields are confined to one of the branes.

We have studied the case when both the colliding energy $\sqrt{s}$
and 5-di\-mensional Planck scale $\bar{M}_5$ are equal to one or
few TeV (and, consequently, $\sqrt{s} \sim \bar{M}_5 \gg \kappa$).
Ultra-high energy region $\sqrt{s} \gg \bar{M}_5$ is also
considered. By taking into account nonzero graviton widths
$\Gamma_n$ (with $n$ being the KK number), we have derived
tree-level analytical expression for $\mathcal{S}(s)$, the
process-independent gravity part of the scattering amplitude (see
our main formulas \eqref{22} and \eqref{24}).

Then we have considered the case when a series of narrow low-mass
graviton resonances is replaced by a continuous mass spectrum.
Both the real and imaginary parts of $\mathcal{S}(s)$ appeared to
differ drastically from those derived without such a replacement.
Thus, one has to conclude that an accurate estimation of the input
sum in $n$ is needed in order to get a correct result, epecially
at $\sqrt{s} \lesssim \bar{M}_5$. The replacement of the summation
in the KK number by integration over graviton mass is well
justified if the graviton mass splitting $\Delta m_{\mathrm{KK}}$
is less than $\Gamma_n$. In its turn, this condition is actually
satisfied only in the trans-Planckian region (when the invariant
energy $\sqrt{s}$ is several times larger than the 5-dimensional
Planck mass).

Zero width approximation (all $\Gamma_n = 0$) has been also
studied. It is shown that some results can not be reproduced in
this approximation, as one can see, for instance, from
Eq.~\eqref{46} which does not admit the limit $\Gamma_n
\rightarrow 0$.

Possible loop corrections to our results originating from the
exchange of $s$-channel gravitons can be estimated in the zero
width approximation as follows. Each new virtual graviton brings
the coupling $1/\Lambda_{\pi}^2$, while a corresponding summation
in $n$ results in the additional dimensional factor $1/\kappa$
(after the substitution $n \rightarrow m_n/(\pi \kappa)$). Using
dimensional arguments, we conclude, for example, that $m$-loop
graviton contribution to a vector field  scattering amplitude,
$\mathcal{M}^{(m)}(s)$, should be proportional to
\begin{equation}\label{loop_corrections}
\mathcal{M}^{(m)}(s) \sim \frac{s^2 M_{cut}^{3m - 1}}{(\kappa
\Lambda_{\pi}^2)^{m+1}} = \frac{s^2 M_{cut}^{3m -
1}}{\bar{M}_5^{3m + 3}} \; ,
\end{equation}
where $M_{cut}$ is the UV cutoff in Feynman integrals, and $m
\geqslant 1$. Thus, multi-loop effects may become dominating at
$M_{cut} \gtrsim \bar{M}_5$. In such a case, an effective operator
analysis will be probably useful (see, for instance,
Ref.~\cite{Giudice:04}). Similar arguments are also applied to the
flat space-time with the large compact ED's, after replacements
$1/\Lambda_{\pi}^2 \rightarrow 1/\bar{M}_{\mathrm{Pl}}^2$,
$1/\kappa \rightarrow R_c^d$. Note that the zero width
approximation is very well justified for this case, since
$\Gamma_n \sim m_n^3/\bar{M_{\mathrm{Pl}}^2}$ (i.e. gravitons are
extremely narrow spin-2 resonances, even for $m_n \sim \sqrt{s}$).

In a general case (nonzero $\Gamma_n$), an additional nontrivial
dependence on the scale $\Lambda_{\pi}$ appears in the RS scheme,
since the graviton widths depend on this mass scale (see
Eq.~\eqref{06}). Both simple dimensional arguments and formulas
like Eq.~\eqref{loop_corrections} are no longer valid.

Our formula~\eqref{22} can be also applied to the scattering
of the brane particles, induced by $t$-channel graviton exchanges.%
\footnote{See also Ref.~\cite{Kisselev:05_2} in which $t$-channel
graviton contribution to the scattering amplitude was studied in a
large curvature scenario of the RS model.}
Let
\begin{equation}\label{kinematical_region}
\frac{\bar{M}_5^3}{\kappa} \gg - t \gg \kappa^2,
\end{equation}
with $t$ being 4-momentum transfer. Then we obtain
from~\eqref{22}:
\begin{equation}\label{t-channel}
\mathcal{S}(t) = - \frac{1}{2 \kappa \bar{M}_5^3} \,
\frac{1}{\tilde{\sigma}} \,
\frac{I_2(\tilde{\sigma})}{I_1(\tilde{\sigma})} \; ,
\end{equation}
where $I_{\nu}(z) = \exp (-i\nu \pi/2) \, J_{\nu} (i z)$ is the
modified Bessel function, and
\begin{equation}\label{parameter_c}
\tilde{\sigma} \simeq  \frac{\sqrt{-t}}{\kappa} - \frac{i \eta}{2}
\, \left( \frac{\sqrt{-t}}{\bar{M}_5}\right)^3 .
\end{equation}
Since $I_2(z)/I_1(z) \rightarrow 1$ at $z \gg 1$ ($- \pi/2 < \arg
z < 3 \pi/2$), we find from \eqref{t-channel} that
\begin{equation}\label{asymptotics}
\mathcal{S}(t) = - \frac{1}{2\bar{M}_5^3 \sqrt{-t}} \;
\end{equation}
in the kinematical region \eqref{kinematical_region}. Note that
$\mathcal{S}(t)$~\eqref{asymptotics} is \emph{pure real} and it
coincides with the imaginary part of $\mathcal{S}(s)$ derived in
the zero width approximation~\eqref{78} up to the replacement $s
\rightarrow -t$. In more general approach, one should sum
KK-charged gravi-Reggeons, i.e. Regge trajectories $\alpha_n(t)$
which are numerated by the KK number
$n$~\cite{Kisselev:04,Kisselev:05}. In such a case, the amplitude
has both real and imaginary parts.

The relations of all cross sections necessary for studying effects
induced by tree-level exchange of the KK gravitons with the
quantities $\mathcal{S}(s)$ and $\mathcal{S}(t)$ can be found in
the Appendix of Ref.~\cite{Giudice:04}.


\setcounter{equation}{0}
\renewcommand{\theequation}{A.\arabic{equation}}

\section*{Appendix A}
\label{app:A}

The sum \eqref{44} can be written as
\begin{equation}\label{A02}
\mathcal{S}(s) = - \frac{i}{\eta \, (\pi \kappa)^4} \,
\sum_{n=1}^{\infty} \frac{1}{n^4 + C^4} \; ,
\end{equation}
where $a$ and $c$ are defined by Eq.~\eqref{10}, and
\begin{equation}\label{A04}
C = e^{- i \pi/8} \, \frac{1}{\pi \kappa} \, \left( \frac{s
\Lambda_{\pi}^2}{\eta}\right)^{1/4}  = x - iy.
\end{equation}
The sum in Eq. \eqref{A02} is of the form~\cite{Prudnikov:I}: 
\begin{equation}\label{A08}
\sum_{n=1}^{\infty} \frac{1}{n^4 + C^4} = \frac{\pi}{2\sqrt{2} \,
C^3} \, \frac{\sinh \sqrt{2} \pi C  + \sin \sqrt{2} \pi C}{\cosh
\sqrt{2} \pi C - \cos \sqrt{2} \pi C} - \frac{1}{2 C^4} \; .
\end{equation}

At large $C$, the main contribution to \eqref{A02} comes from $n
\sim |C|$. The disregard of term $b x_n^2$ in \eqref{08} is
justified if it is much less than $c$, that results in the
inequality $b \pi^2 |C|^2 \ll c$, or, equivalently, $\sqrt{\eta s}
\gg \Lambda_{\pi}$~\eqref{42}.

By using formulas
\begin{align}
\sin (x-iy) &= \sin x \cosh y - i \cos x \sinh y,
\nonumber \\
\cos  (x-iy) &= \cos x \cosh y + i  \sin x \sinh y,
\label{A10} \\
\intertext{and}
\sinh (x-iy) &= \cos y \sinh x - i \sin y \cosh x,
\nonumber \\
\cosh (x-iy) &= \cos y \cosh x + i \sin y \sinh x , \label{A12}
\end{align}
and taking into account that $\cosh x \simeq \sinh x \gg \cosh y
\simeq \sinh y$ is valid at $x = \cot (\pi/8) \, y \simeq 2.4 \, y
\gg 1$, we obtain formula~\eqref{46}.

\setcounter{equation}{0}
\renewcommand{\theequation}{B.\arabic{equation}}

\section*{Appendix B}
\label{app:B}

At small ratio $\kappa/\sqrt{s}$, we get from Eq.~\eqref{50} that
\begin{equation}\label{B02}
\mathrm{Re} \, \mathcal{S} _0(s) = \frac{1}{\pi \sqrt{s}
\bar{M}_5^3} \left[ \mathcal{I} - \frac{\pi
\kappa}{\sqrt{s}}\right],
\end{equation}
where
\begin{equation}\label{B04}
\mathcal{I} =  \int\limits_0^{\infty} \! dz \, \frac{1 - z^2}{(1 -
z^2)^2 + \delta z^8} \; ,
\end{equation}
with $\delta$ being defined by Eq.~\eqref{58}.

It is convenient to divide our integral \eqref{B04} into two
parts:
\begin{equation}\label{B06}
\mathcal{I} = \mathcal{I}_1 + \mathcal{I}_2 \; .
\end{equation}
Here
\begin{align}
\mathcal{I}_1 &= \int\limits_0^1 \! du \, \left[ \frac{u(2 -
u)}{u^2(2 - u)^2 + \delta (1 - u)^8} - \frac{u(2 + u)}{u^2(2 +
u)^2 + \delta (1 + u)^8} \right],
\label{B08} \\
\intertext{and} \mathcal{I}_2 &= - \int\limits_1^{\infty} \! du \,
\frac{u(2 + u)}{u^2(2 + u)^2 + \delta (1 + u)^8} \; .
\label{B10}
\end{align}

Up to higher powers of $\delta$, the integrals \eqref{B08} and
\eqref{B10} are equal to
\begin{equation}\label{B12}
\mathcal{I}_1 = \frac{1}{2} \ln 3  + \frac{3 \pi}{4} \,
\sqrt{\delta} \; , \quad \mathcal{I}_2 = - \frac{1}{2} \ln 3 +
\frac{\pi}{2\sqrt{2}} \, \sqrt[4]{\delta} \; ,
\end{equation}
and we obtain formula \eqref{60} of the main text.

As for the imaginary part, it is given by
\begin{equation}\label{B14}
\mathrm{Im} \, \mathcal{S} _0(s) = - \frac{\eta \sqrt{s}}{\pi
\kappa \Lambda_{\pi}^4} \; \mathcal{J},
\end{equation}
where
\begin{equation}\label{B16}
\mathcal{J} =  \int\limits_0^{\infty} \! dz \, \frac{z^4}{(1 -
z^2)^2 + \delta z^8} \; .
\end{equation}

As in the previous case, we divide the integral in \eqref{B16}
into two parts:
\begin{equation}\label{B18}
\mathcal{J} = \mathcal{J}_1 + \mathcal{J}_2,
\end{equation}
with
\begin{align}
\mathcal{J}_1 &= \int\limits_0^1 \! du \, \left[ \frac{(1 -
u)^4}{u^2(2 - u)^2 + \delta (1 - u)^8} + \frac{(1 + u)^4}{u^2(2 +
u)^2 + \delta (1 + u)^8} \right],
\label{B20} \\
\intertext{and} \mathcal{J}_2 &= \int\limits_1^{\infty} \! du \,
\frac{(1 + u)^4}{u^2(2 + u)^2 + \delta (1 + u)^8} \; .
\label{B22}
\end{align}

The leading parts of the integrals \eqref{B20} and \eqref{B22} are
equal to
\begin{equation}\label{B24}
\mathcal{J}_1 = \frac{\pi}{2 \sqrt{\delta}} \; , \quad
\mathcal{J}_2 = \frac{\pi}{2\sqrt{2}} \,
\frac{1}{\sqrt[4]{\delta}} \; .
\end{equation}
Thus, we get
\begin{equation}\label{B26}
\mathcal{J} \simeq \frac{\pi}{2 \eta} \, \frac{\Lambda_{\pi}^2}{s}
\; ,
\end{equation}
that results in formula~\eqref{62} in the main text.

\setcounter{equation}{0}
\renewcommand{\theequation}{C.\arabic{equation}}

\section*{Appendix C}
\label{app:C}

Let us consider the sum
\begin{equation}\label{C02}
\mathcal{K} = \sum_{n=1}^{\infty} \frac{1}{\displaystyle u^2 - ( n
+ v)^2} \; ,
\end{equation}
with
\begin{equation}\label{C04}
u = \frac{\sqrt{s}}{\pi \kappa} + i0  \; , \quad v = \frac{1}{4}
\end{equation}
(remember that $\sqrt{s} \gg \kappa$).

The infinite sum~\eqref{C02} can be found in Ref.~\cite{Prudnikov:I}: 
\begin{equation}\label{C06}
\mathcal{K} = \frac{1}{2u} \left[ \Psi(v - u) - \Psi(v + u)\right]
- \frac{1}{u^2 - v^2} \; ,
\end{equation}
where $\Psi(z)$ is the psi-function. Then one can apply the
formula 
\begin{equation}\label{C08}
\Psi(-z) = \Psi(z) + \frac{1}{z} + \pi \cot(\pi z) \; ,
\end{equation}
and use the asymptotic behavior of the function $\Psi(z)$ at large
$z$ ($|\arg z| <
\pi$), 
\begin{equation}\label{C10}
\Psi(z) \big|_{|z| \gg 1} = \ln z - \frac{1}{2z} +
\mathrm{O}(z^{-2})
\end{equation}
(both formulas are taken from \cite{Erdelyi:I}).

As a result, the asymptotics of $\mathcal{K}$ looks like
\begin{equation}\label{C12}
\mathcal{K} \, \big|_{|u| \gg 1} = \frac{\pi}{2u} \, \cot[\pi (u -
v)] + \mathrm{O}(u^{-2}) \; ,
\end{equation}
and we come to Eq.~\eqref{72} presented in the text.


\end{document}